# Enhancement of spin-orbit torque efficiency by tailoring interfacial spin-orbit coupling in Pt-based magnetic multilayers

Wenqiang Wang(王文强)[1], Kaiyuan Zhou(周恺元)[1], Xiang Zhan(战翔)[1], Zui Tao(陶醉)[1], Qingwei Fu(付清为)[1], Like Liang(梁力克)[1], Zishuang Li(李子爽)[1], Lina Chen(陈丽娜)[1,2,*], Chunjie Yan(晏春杰)[1], Haotian Li(李浩天)[1], Tiejun Zhou(周铁军)[3,*], and Ronghua Liu(刘荣华)[1,*]

[1] National Laboratory of Solid State Microstructures, School of Physics and Collaborative Innovation Center of Advanced Microstructures, Nanjing University, Nanjing 210093, China
[2] New Energy Technology Engineering Laboratory of Jiangsu Provence and School of Science, Nanjing University of Posts and Telecommunications, Nanjing 210023, China
[3] Centre for Integrated Spintronic Devices, School of Electronics and Information, Hangzhou Dianzi University, Hangzhou 310018, China

[*] E-mail: rhliu@nju.edu.cn; chenlina@njupt.edu.cn and tjzhou@hdu.edu.cn

We study inserting Co layer thickness-dependent spin transport and spin-orbit torques (SOTs) in the Pt/Co/Py trilayers by spin-torque ferromagnetic resonance. The interfacial perpendicular magnetic anisotropy energy density ($K_s = 2.7$ erg/cm$^2$), which is dominated by interfacial spin-orbit coupling (ISOC) in the Pt/Co interface, total effective spin-mixing conductance ($G_{\text{eff,tot}}^{\uparrow\downarrow} = 0.42 \times 10^{15} \Omega^{-1}\text{m}^{-2}$) and two-magnon scattering ($\beta_{\text{TMS}} = 0.46$ nm$^2$) are first characterized, and the damping-like torque ($\xi_{\text{DL}} = 0.103$) and field-like torque ($\xi_{\text{FL}} = -0.017$) efficiencies are also calculated quantitatively by varying the thickness of the inserting Co layer. The significant enhancement of $\xi_{\text{DL}}$ and $\xi_{\text{FL}}$ in Pt/Co/Py than Pt/Py bilayer system originates from the interfacial Rashba-Edelstein effect due to the strong ISOC between Co-*3d* and Pt-*5d* orbitals at the Pt/Co interface. Additionally, we find a considerable out-of-plane spin polarization SOT, which is ascribed to the spin anomalous Hall effect and possible spin precession effect due to IPMA-induced perpendicular magnetization at the Pt/Co interface. Our results demonstrate that the ISOC of the Pt/Co interface plays a vital role in spin transport and SOTs-generation. Our finds offer an alternative approach to improve the conventional SOTs efficiencies and generate unconventional SOTs with out-of-plane spin polarization to develop low power Pt-based spintronic via tailoring the Pt/FM interface.

## 1. Introduction

Spin-orbit torques [1-8] in heavy metal/ferromagnet (HM/FM) systems have become a powerful approach to achieving a pure current control of magnetization switch for building an energy-efficient SOT-magnetoresistive random-access memory (SOT-MRAM)[9] and excitation of coherent spin waves for magnon-based logic devices[10,11] or spin synchronization-based neuromorphic computing[12-14].

The generation of SOTs originates from the orbital angular momentum transferred from the lattice to the spin system due to the spin-orbit interaction in the HM with strong bulk spin-orbit coupling (SOC) or/and at the HM/FM interface with strong interfacial spin-orbit coupling. The former is generally described as the spin Hall effect (SHE), and the latter is commonly known as the interfacial Rashba-Edelstein effect (IREE)[15]. As shown in Fig. 1(a), SHE/IREE-generated in-plane (IP) transverse polarization $\boldsymbol{\sigma_y}$ spin currents exert two types of SOTs on the magnetization $\boldsymbol{m}$ of the adjacent FM layer: one is the IP damping-like (DL) torque[1,16] $\boldsymbol{\tau}_{\mathrm{DL}} = \tau_{\mathrm{DL}} \boldsymbol{m} \times (\boldsymbol{\sigma} \times \boldsymbol{m})$; and the other is the out-of-plane (OP) field-like (FL) torque[17] $\boldsymbol{\tau}_{\mathrm{FL}} = \tau_{\mathrm{FL}}(\boldsymbol{\sigma} \times \boldsymbol{m})$. Previously many theoretical studies have revealed that the DL torque originates predominantly from the SHE[18], while the FL torque is dominated by IREE[15]. However, many recent FM thickness- and interface-dependent SOTs efficiency experiments found that the IREE can also significantly contribute to the DL torque with a value comparable to SHE-contribution in some HM/FM systems with strong ISOC. In addition, the interface-related spin memory loss (SML) and spin flow back also play an essential role in the effective SOTs efficiencies. For instance, the SOTs efficiencies can be significantly enhanced by improving interfacial spin transmission efficiency and enhancing IREE via interface engineering, such as inserting ultrathin nonmagnetic metal layer (e.g., Hf[19], Mo[20], and Cu[21,22]) between the HM and FM layers, oxygen-induced interface orbital hybridization[23], alloy [24]and interfacial $H^+$ and $O^{2-}$ ion manipulations[25]. Very recently, inserting a magnetic spacer[26,27] between the FM and HM layers has also been proposed as a promising method to improve the effective SOTs efficiencies by selecting suitable magnetic materials to increase ISOC and/or enhance interfacial spin transparency ($T_{int}$)[26]. However, the detailed characterization of the ISOC-related interfacial magnetic properties, e.g., interfacial perpendicular magnetic anisotropy (IPMA) energy density[19], $T_{int}$, two-magnon scattering (TMS) coefficient[28], the conventional SHE/IREE-induced damping-like torque efficiency ($\xi_{DL}$) and field-like torque efficiency ($\xi_{FL}$) and possible unconventional SOTs of spin currents with OP spin polarization in Pt-based multilayer systems with a strong ISOC still remains a few so far.

In this letter, we study the SOTs efficiencies in the Pt/Co (*t*)/Py trilayers with a strong ISOC at the Pt/Co interface by spin-torque ferromagnetic resonance (ST-FMR) technique and find a significant enhancement of the $\xi_{DL} = 0.103$ and $\xi_{FL} = -0.017$ in Pt/Co/Py system with an inserting Co layer compared to $\xi_{DL} = 0.051$ and $\xi_{FL} = -0.002$ in the Pt/Py bilayer. The enhancement of SOT efficiencies is related to the IREE-induced additional SOTs at a robust ISOC Pt/Co interface by inserting an ultrathin Co layer between Pt and Py. In addition, these Pt/Co/Py trilayers also exhibit considerable OP spin-polarized SOTs, revealed by the IP angular dependence ST-FMR spectra. The results suggest that the ISOC of the Pt/Co interface plays a vital role in spin transport and SOTs-generation and can be an efficient approach to promote the conventional SOTs efficiency and generate unconventional SOTs for the development of non-volatility, low power, higher speed, and higher endurance spintronic devices.

## 2. Experiments

The ST-FMR devices consist of the stack structure: Pt(5)/Co(*t* = 0, 0.2, 0.3, 0.4, 0.5, 0.75, 1, 1.5, and 2)/Py(4), which are deposited on annealed $Al_2O_3$ substrate with (0001) orientation by d.c magnetron sputtering at room temperature with Ar sputtering gas pressure of 5.1 mTorr and background base pressure of $2 \times 10^{-8}$ Torr. The film thickness in parentheses is in nm. A 2-nm-thick MgO is adopted to protect the multilayers from oxidation in air. The films are patterned into a 5 um × 8 um rectangle stripe with two top electrodes of Au (80) for ST-FMR measurement using the combination of photolithography, electron beam lithography, and ion milling.

## 3. Results and discussion

Figure 1(a) shows the ST-FMR measurement setup[29], where a radio-frequency (RF) current ($I_{RF}$) is applied along the longitudinal direction of the stripe by connecting a signal generator, and a dc mixing voltage ($V_{mix}$), generated from rectification among the RF current, SOTs and Oersted field-driven oscillating resistance due to anisotropic magnetoresistance (AMR), is recorded by a lock-in while sweeping an IP external field $H$. Figure 1(b) shows the representative ST-FMR spectra of Pt(5)/Co(0.4)/Py(4) sample with excitation frequency from 6 to 10 GHz, IP angle $\varphi = 30°$. The obtained $V_{mix}$ can be well fitted with a Lorentzian function [1,23]:

$$V_{mix} = V_s \frac{\Delta H^2}{[(H-H_{res})^2+\Delta H^2]} + V_a \frac{\Delta H(H-H_{res})}{[(H-H_{res})^2+\Delta H^2]} \tag{1}$$

where $V_s$, $V_a$, $\Delta H$ and $H_{res}$ are the magnitude of the symmetric ($V_s$) and antisymmetric ($V_a$) Lorentzian components, the linewidth, and resonance field, respectively. The representative $V_{mix}$ data of the Pt (5)/Co (0.4)/Py(4) with $f$ = 6 GHz and its fitting curves are illustrated in Fig. 1(c).

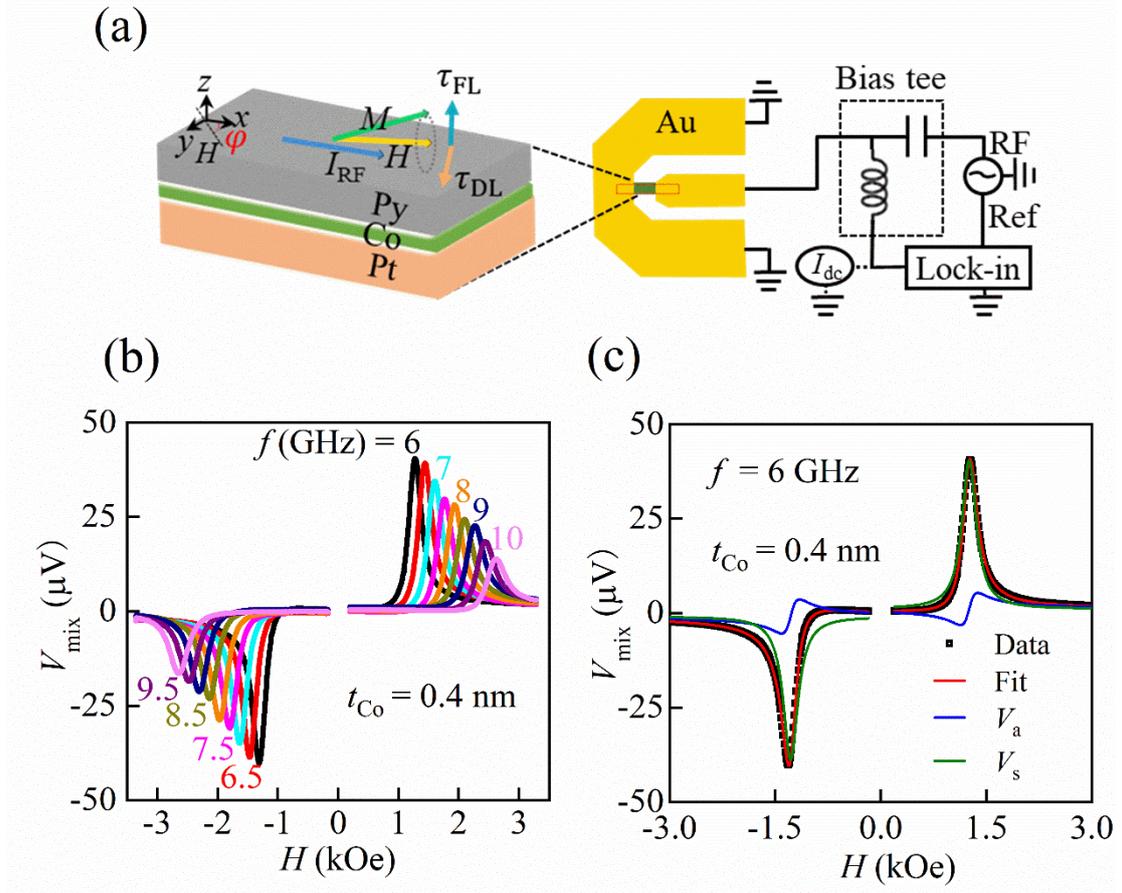

Fig.1. ST-FMR spectra of Pt/Co/Py. (a) Left: Illustration of the stack structure of multilayer, coordinate system, and SOTs-induced magnetization dynamics in the ST-FMR measurement. Right: schematic diagram of the ST-FMR setup. (b) ST-FMR spectra of the $V_{\text{mix}}$ of the Pt (5)/Co(0.4)/Py (4) sample for frequency $f$ between 6 and 10 GHz increased in 0.5 GHz step at an angle $\varphi = 30°$ between the magnetic field and current direction. (c) Representative ST-FMR spectrum obtained at $f$ = 6 GHz and its fitting curves using Eq. (1). $V_s$ and $V_a$ correspond to the symmetric and antisymmetric Lorentzian components, respectively.

Before discussing the spin torques efficiencies, we first characterize the interfacial properties of the Pt/Co/Py trilayers by quantifying the magnetic anisotropy and effective spin-mixing conductance ($G_{\text{eff,tot}}^{\uparrow\downarrow}$) related to the ISOC. The total effective magnetic anisotropy energy density $K_{\text{eff}}$ (erg/cm$^3$) consists of the volume contribution $K_v$ and the interfacial $K_s = K_s^{\text{Pt/Co}} + K_s^{\text{Co/Py}}$ with the following relation[30,31]:

$$K_{\text{eff}}t_{\text{FM}}^{\text{tot}} = K_{\text{v}}t_{\text{FM}}^{\text{tot}} + K_{\text{s}} \qquad (2)$$

The effective demagnetization field $4\pi M_{\text{eff}} = 4\pi M_{\text{s}} - K_{\text{eff}}/2\pi M_{\text{s}}$ for all studied ST-FMR devices can be obtained by fitting the experimental results of $f$ vs. $H_{\text{res}}$ [Fig. 2 (a)] using the Kittel formula[23]: $f = (\gamma/2\pi)\sqrt{H_{\text{res}}(H_{\text{res}} + 4\pi M_{\text{eff}})}$, where $\gamma/2\pi$ is the gyromagnetic ratio. Figure 2(b) shows the $4\pi M_{\text{eff}}$ as a function of the inserting Co layer thickness $t_{\text{Co}}$. To quantitatively extract the value of $K_{\text{eff}}$, we also measure the saturation magnetization $M_{\text{s}}$ of all samples by vibrating sample magnetometry (VSM), as shown in Fig. 2(c). The $M_{\text{s}}$ of all samples can be well fitted by the formula $M_{\text{s}} = \left(M_{\text{s}}^{\text{Py}}t_{\text{Py}} + M_{\text{s}}^{\text{Co}}t_{\text{Co}}\right)/(t_{\text{Py}} + t_{\text{Co}})$, with two reasonable parameters $M_{\text{s}}^{\text{Py}} = 707$ emu/cm$^3$ and $M_{\text{s}}^{\text{Co}} = 1125$ emu/cm$^3$, consistent with previously reported values[32,33] $M_{\text{s}}^{\text{Py}} = 697$ emu/cm$^3$ and $M_{\text{s}}^{\text{Co}} = 1084$ emu/cm$^3$. Therefore, we can determine the $K_{\text{eff}}$ from the effective demagnetization field. Figure 2(d) shows that $K_{\text{eff}}t_{\text{FM}}^{\text{tot}}$ for all inserting Co samples exhibits a linear dependence on $t_{\text{FM}}^{\text{tot}}$ except for the Pt/Py bilayer sample, indicating that the interfacial $K_{\text{s}}$ is dominated by the Pt/Co and Co/Py interface. From the intercept of the linear fitting with Eq. (2) at $t_{\text{FM}}^{\text{tot}} = 0$, we estimate $K_{\text{s}} = 2.7$ erg/cm$^2$, comparable with previous reports value[33,34]. Meanwhile, the effective interfacial PMA field ($H_\perp$) $H_\perp = 4\pi M_{\text{s}} - 4\pi M_{\text{eff}}$ is also illustrated in the inset of Fig. 2(d), indicating that inserting a thin Co layer between Pt and Py can lead to a large effective OP magnetic anisotropy. The reason is that the 5$d$ - Pt with a strong SOC can modify the perpendicular orbital moments in an adjacent 3$d$ - Co layer via strong interfacial 3$d$ - 5$d$ hybridization[35,36].

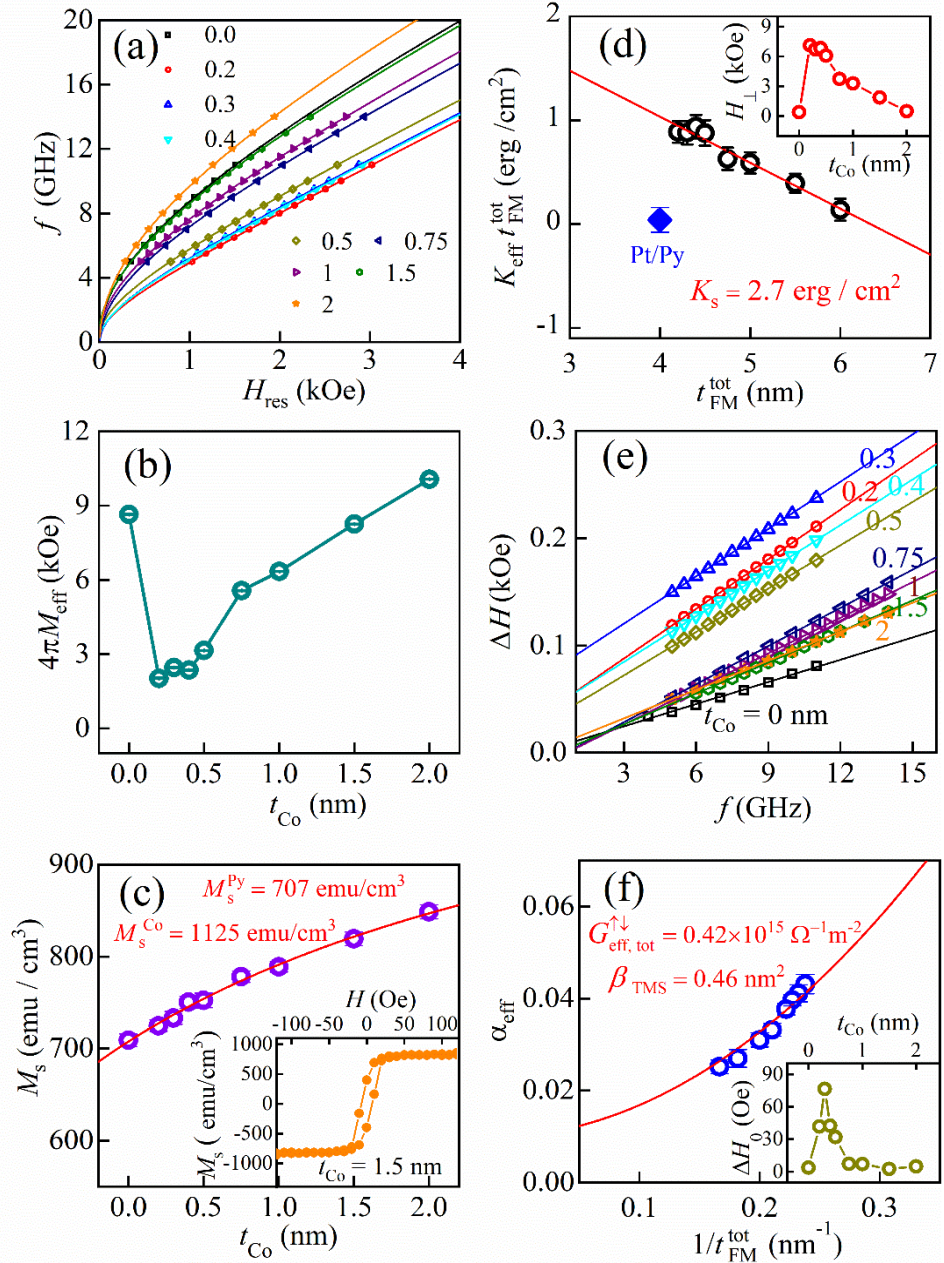

Fig. 2. Thickness-dependent magnetic properties of Py(5)/Co($t_{Co}$ = 0 - 2 nm)/Py(4) system. (a) Dispersion relation curves between $f$ and $H_{res}$ (symbols) and the corresponding Kittel fittings (solid curves). (b) - (d) Dependence of the effective demagnetization field $4\pi M_{eff}$ (b), and the saturation magnetization $M_s$ (c) and the $H_\perp$ (Inset in (d)) on the inserting Co layer thickness $t_{Co}$, and the effective magnetic anisotropy energy in terms of $K_{eff} t_{FM}^{tot}$ (d) on the total FM layer thickness $t_{FM}^{tot}$. Inset in (c): the representative magnetization curve of $t_{Co}$ = 1.5 nm sample. (e) Linewidth $\Delta H$ versus resonance frequency $f$. The solid line is the linear fitting. (f) Damping $\alpha_{eff}$

as a function of $1/t_{\text{FM}}^{\text{tot}}$ (symbols). Inset: dependence of the inhomogeneous line broadening $\Delta H_0$ on the inserting Co layer thickness $t_{\text{Co}}$.

In a magnetic heterostructure HM/FM system, the $G_{\text{eff,tot}}^{\uparrow\downarrow}$ is a key parameter of the indication of spin transmission and SOTs efficiencies, e.g., the $T_{\text{int}}$ is proportional to $G_{\text{eff,tot}}^{\uparrow\downarrow}$, and the $\xi_{\text{DL}}$ depends on $G_{\text{eff,tot}}^{\uparrow\downarrow}$ in terms of $\xi_{\text{DL}} = T_{\text{int}}\theta_{\text{SH}}$, where $\theta_{\text{SH}}$ is the intrinsic charge-to-spin convert ratio or spin Hall angle due to SHE/IREE. Therefore, we can get some helpful information about the interface-related spin transparency and spin-mixing conductance from the magnetic damping. The inserting Co layer thickness-dependent effective damping ($\alpha_{\text{eff}}$) can be determined by fitting the experimental results of $f$ vs. linewidth $\Delta H$ [Fig. 2(e)] using[37,38] $\Delta H = (2\pi/\gamma)\alpha_{\text{eff}} f + \Delta H_0$, where $\Delta H_0$ is the inhomogeneous linewidth. $\Delta H_0$ exhibits a significant enhancement for the samples by inserting Co thickness $t_{\text{Co}} = 0.2 - 0.5$ nm [Inset of Fig. 2(f)], which is related to the inhomogeneity of interfacial magnetic anisotropy and magnetic properties caused by the discussed ISOC of Pt/Co above. The extracted $\alpha_{\text{eff}}$ as a function of the total thickness of the FM layer is shown in Fig. 2 (f). As we know, besides the FM layer thickness-independent intrinsic Gilbert damping ($\alpha_{\text{int}}$), the $\alpha_{\text{eff}}$ contains additional two main contributions: $G_{\text{eff,tot}}^{\uparrow\downarrow}$ related to spin pumping and TMS due to ISOC and magnetic defects at interfaces. The total $\alpha_{\text{eff}}$ is given approximately [28,39]

$$\alpha_{\text{eff}} = \alpha_{\text{int}} + G_{\text{eff,tot}}^{\uparrow\downarrow}\frac{g\mu_B h}{4\pi M_s e^2}\frac{1}{t_{\text{FM}}^{\text{tot}}} + \beta_{\text{TMS}}\frac{1}{t_{\text{FM}}^{\text{tot}-2}} \qquad (3)$$

Where $g$ is the Lande factor, $\mu_B$ is the Bohr magnetron, and $h$ is the Planck's constant. The second term is related to the spin currents loss via spin pumping into the Pt layer and being absorbed due to SML at the Pt/Co and Co/Py interfaces[39], and the third term is the contribution from the TMS process, where the TMS coefficient $\beta_{\text{TMS}}$ depends on both $(2 K_s/M_s)^2$ and the density of magnetic defect at the interfaces. Fitting the dependence of $\alpha_{\text{eff}}$ on $1/t_{\text{FM}}^{\text{tot}}$ in Fig. 2 (f) by using Eq. (3), we determine the $\alpha_{\text{int}} = 0.010$ of FM layer and the $G_{\text{eff,tot}}^{\uparrow\downarrow} = 0.42 \times 10^{15}\Omega^{-1}m^{-2}$, which is comparable with the previously reported value $G_{\text{eff,tot}}^{\uparrow\downarrow} = 0.31 \times 10^{15} \Omega^{-1} \text{m}^{-2}$ of the Pt/Co bilayer[28,39], suggesting the Pt/Co interface dominates the interfacial SOTs efficiencies and $T_{\text{int}}$. Additionally, we also find that the TMS is nonnegligible and the fitting parameter $\beta_{\text{TMS}} = 0.46 \text{ nm}^2$ indicates that the Pt/Co/Py trilayer has a much stronger TMS than the Pt/Py bilayer due to the strong ISOC-induced IPMA and interfacial scattering.

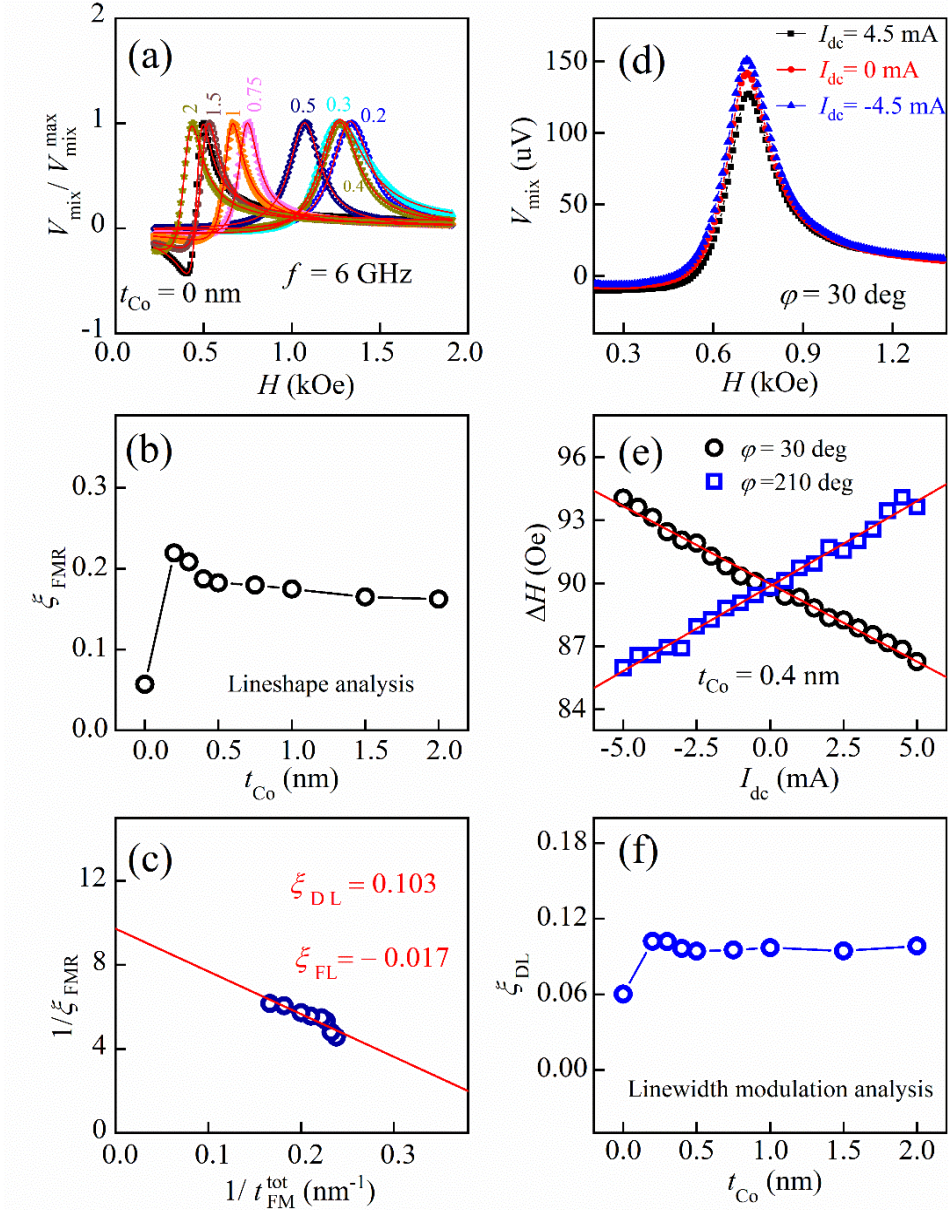

Fig. 3. Thickness-dependent $\xi_{\text{DL}}$ and $\xi_{\text{FL}}$. (a) The normalized ST-FMR spectra voltage of all studied samples with $t_{\text{Co}} = 0 - 2$ nm at $f = 6$ GHz (symbols) and the fitting results using Eq. (1) (solid lines). (b) The $\xi_{FMR}$ of all films as a function of the interlayer Co thickness evaluated by the ratio $V_s/V_a$. (c) The relation of $1/\xi_{FMR}$ vs. $1/t_{FM}^{tot}$ and its linear fitting using Eq. (5) in the text. (d) ST-FMR spectra at different dc bias currents $I_{dc}$ with an IP angle $\varphi = 30°$ for the $t_{Co} = 0.4$ nm sample. (e) Linewidth as a function of applied dc bias current $I_{dc}$ in the case of IP angle $\varphi = 30°$ (black circles) and $\varphi = 210°$ (red squares) for the $t_{Co} = 0.4$ nm sample. The solid red lines are the linear fittings. (f) The inserting Co layer thickness $t_{Co}$ dependence of $\xi_{DL}$ determined by the LWM method.

To further examine how the above Pt/Co interfacial characteristics of $K_s$, $G_{eff,tot}^{\uparrow\downarrow}$ and the coefficient $\beta_{TMS}$ are reflected in the $\xi_{DL}$ and $\xi_{FL}$, we adopt two methods of lineshape analysis ($V_s/V_a$) and linewidth modulation (LWM) to quantify the SOTs via the ST-FMR spectra. Based on the previous reports[1,23] of the HM/FM bilayer system with the conventional SHE/IREE, the $V_s$ and $V_a$ components originate from the OP DL effective field $H_{DL}$ and IP Oersted field $H_{Oe}$ and/or FL effective field $H_{FL}$, respectively. Therefore, the spin-torque efficiency ($\xi_{FMR}$) can be estimated by the $V_s/V_a$ ratio as[33,34,40]

$$\xi_{FMR} = \frac{V_s}{V_a} \frac{e 4\pi M_s t_{FM}^{tot} t_{HM}}{\hbar} \sqrt{1 + 4\pi M_{eff}/H_{res}} \qquad (4)$$

where $t_{HM}$ and $t_{FM}^{tot}$ represent thicknesses of the Pt layer and the total FM layer, respectively. $\xi_{FMR}$ depends on $t_{FM}^{tot}$ because $H_{FL}$ is inversely proportional to the FM layer thickness in terms of $H_{FL} \propto \xi_{FL}/t_{FM}^{tot}$ and $H_{Oe}$ is independent of $t_{FM}^{tot}$. Furthermore, $\xi_{FMR}$ can be divided into the $\xi_{DL}$ and $\xi_{FL}$ as the following formula:

$$\frac{1}{\xi_{FMR}} = \frac{1}{\xi_{DL}}\left(1 + \frac{\hbar}{e}\frac{\xi_{FL}}{4\pi M_s t_{FM}^{tot} t_{HM}}\right) \qquad (5)$$

Where $e$ is the electronic charge, $\hbar$ is the reduced Planck's constant. To disentangle SOTs efficiencies, we need to get thickness-dependent $\xi_{FMR}$ by varying the thickness of the inserting Co layer. Figure 3(a) shows the normalized ST-FMR spectra voltage of all studied samples with $t_{Co} = 0 - 2$ nm at $f = 6$ GHz. The $V_s$ and $V_a$ components can be determined by fitting ST-FMR spectra using a Lorentzian function Eq. (1). According to Eq. (4), we calculate the $\xi_{FMR}$ of all samples as a function of the intercalation Co thickness $t_{Co}$, as shown in Fig. 3(b). Then, the $\xi_{DL} = 0.103$ and $\xi_{FL} = -0.017$ are determined by using a linear function Eq. (5) to fit $1/\xi_{FMR}$ vs. $1/t_{FM}^{tot}$ data [Fig. 3(c)]. The obtained values are consistent with the previous reports in the pure Pt/Co bilayer systems[33].

To further confirm the validity of the above determined $\xi_{DL}$, we also adopt an alternative approach to quantify $\xi_{DL}$ based on the LWM by dc current-induced DL SOT acting on the FM layer[23]. $\xi_{DL}$ can be extracted through dc current-dependent linewidth $\Delta H$ measurements by the following formula[1,23]:

$$\xi_{DL} = \frac{\Delta H/I_{dc}}{\frac{2\pi f}{\gamma}\frac{sin\varphi}{(H_{res}+2\pi M_{eff})4\pi M_s t_{FM}^{tot}}\frac{\hbar}{2e}} \frac{R_{FM}+R_{HM}}{R_{FM}} A_c \quad (6)$$

Where $R_{FM}$ and $R_{HM}$ are the resistance of the total FM and Pt layers, respectively, and $A_c$ is the cross-sectional area of the Pt layer. Figure 3(d) shows the representative ST-FMR spectra at different dc bias currents $I_{dc}$ at an IP angle $\varphi = 30°$. The extracted current-dependent linewidth $\Delta H$ at two IP angles $\varphi = 30°$ and $210°$ for $t_{Co} = 0.4$ nm sample at $f = 4$ GHz were shown in Fig. 3(e). The obtained $\xi_{DL}$ using Eq. (6) exhibits a nearly constant value of $\xi_{DL} \approx 0.10$ for all inserting Co samples [Fig. 3(f)], closing to 0.103 obtained by the above $V_s/V_a$ method. These $\xi_{DL} = 0.103$ and $\xi_{FL} = -0.017$ are significantly larger than $\xi_{DL} = 0.051$ and $\xi_{FL} = -0.002$ in the Pt/Py bilayer[1], indicating that the ISOC at the interface Pt/Co plays a vital role in both generating SOTs and spin transmission discussed above.

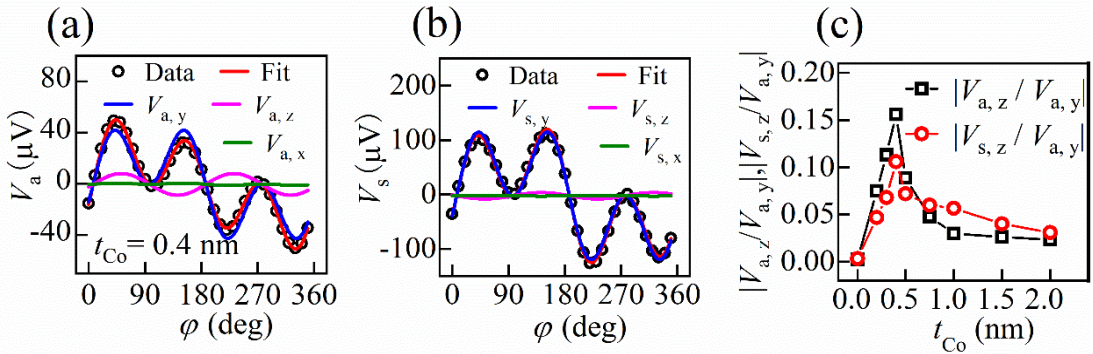

Fig. 4. Unconventional SOTs with OP spin polarization in Pt/Co ($t_{Co}$ = 0.2 - 2 nm)/Py.

(a)-(b) IP angular dependent $V_a$ (a) and $V_s$ (b) of ST-FMR voltage signal for a representative Pt (5)/Co (0.4)/Py(4) sample at $f = 6$ GHz can be well fitted by the combination of three terms: $V_y\sin2\varphi\cos\varphi$ (blue line), $V_z\sin2\varphi$ (pink line) and $V_x\sin2\varphi\sin\varphi$ (green line). (c) The absolute value of $V_{a,z}/V_{a,y}$ and $V_{s,z}/V_{a,y}$ as a function of $t_{Co}$.

Since the ST-FMR voltage signal $V_{mix}$ of our studied microscale device is mainly contributed by the rectification of $I_{RF}$ and $I_{RF}$-induced periodic magnetoresistance due to the AMR effect, and with a negligible inverse SHE voltage due to spin pumping, the IP angular dependence of the $V_a$ and $V_s$ components of $V_{mix}$ is related to the combination of the angular dependences of the AMR and the generated spin currents and the current-induced Oersted field. The AMR exhibits the sinusoidal dependence on magnetization with a period of 180º, $R_{AMR} \propto \sin2\varphi$. Considering the generated spin currents with all possible spin polarizations, we can derive the general formula for describing IP angular-dependent $V_a$ and $V_s$ of ST-FMR voltage as follows [41,42]:

$$V_a = (V_{a,y}\cos\varphi + V_{a,z} + V_{a,x}\sin\varphi)\sin2\varphi \quad (7)$$

$$V_s = (V_{s,y}\cos\varphi + V_{s,z} + V_{s,x}\sin\varphi)\sin2\varphi \quad (8)$$

The $V_{a,y}$ corresponds to the OP torque arising from current-induced IP Oersted field and the effective field due to the generated spin currents with spin polarization ($\boldsymbol{\sigma}_y$) via SHE/IREE, and $V_{s,y}$ is attributed to the SHE/IREE-induced IP DL torque. The $V_{a,z}$ and $V_{s,z}$ originate from the OP DL torque $\tau_{\perp,DL}$ and the IP FL torque $\tau_{\parallel,FL}$ generated by the spin currents with OP spin polarization ($\boldsymbol{\sigma}_z$), respectively. The $V_{a,x}$ and $V_{s,x}$ are correlated with the OP FL torque and IP DL torque due to the spin currents with spin polarization ($\boldsymbol{\sigma}_x$) along the charge current direction (x-direction).

To explore the possible existence of unconventional SOTs arising from the spin currents with $\boldsymbol{\sigma}_z$ and $\boldsymbol{\sigma}_x$ except for conventional $\boldsymbol{\sigma}_y$, we perform the IP angular dependence of the ST-FMR spectra. Figures 4(a) and 4(b) show the IP angular dependent $V_a$ and $V_s$ and their fitting curves using Eq. (7) and Eq. (8) for the representative sample of Pt (5)/Co (0.4)/Py (4) under $f = 6$ GHz. The fitting parameters $V_{a,x}$ and $V_{s,x}$ have the negligible small value compared to $V_{a,y}$, indicating that the generated spin currents with $\sigma_x$ are negligible. However, there exists a considerable value for $V_{a,z}$ and $V_{s,z}$, which can be used to quantify the strength of $\tau_{\perp,DL}$ and $\tau_{\parallel,FL}$ by directly comparing $V_{a,z}$ and $V_{s,z}$ to $V_{a,y}$ because $V_{a,y}$ is mainly proportional to $I_{RF}$-induced Oersted field. Therefore, $V_{a,z}/V_{a,y}$ and $V_{s,z}/V_{a,y}$ represent the strengths of the OP DL and IP FL torques exerted by the generated spin currents with $\sigma_z$. Figure 4(c) shows that $|V_{a,z}/V_{a,y}|$ and $|V_{s,z}/V_{a,y}|$ as a function of the inserting Co layer thickness for all studied samples exhibit a maximum at $t_{Co} = 0.4$ nm. At the same thickness, the $H_\perp$ also exhibits the maximum. The generated spin currents with the OP spin polarization are proportional to the $H_\perp$, suggesting that the strong 3d-Co and 5d-Pt orbital hybridization at the Pt/Co interface can generate a considerable OP DL torque $\tau_{\perp,DL}$ to facilitate current-induced magnetization switching in the HM/FM systems without needing an assistant field[43]. However, in the previous theory reports[44,45], several proposed possible mechanisms, e.g., spin swapping[44], spin-orbit precession[46], and/or spin anomalous Hall effect[47,48], can generate these spin currents with $\sigma_z$, warranting further theoretical and experimental studies of the behind mechanism of interface dependence of this $\sigma_z$-spin currents.

## 4. Conclusion

In summary, we systematically investigate the ISOC phenomena (IPMA coefficient $G_{\text{eff,tot}}^{\uparrow\downarrow}$, TMS coefficient, and SOTs efficiencies) in Pt/Co/Py trilayers system by performing the Co interlayer thickness dependence and IP angular dependence of ST-FMR spectra measurements. Compare to SHE-induced $\xi_{DL} = 0.051$ and $\xi_{FL} = -0.002$ in the Pt/Py bilayer, we experimentally demonstrate that the Pt/Co/Py trilayer has a significant enhancement in SHE-generated IP transverse polarization $\sigma_y$ spin current torques $\xi_{DL} = 0.103$ and $\xi_{FL} = -0.017$. An 80% enhancement in the $\xi_{DL}$ is primarily contributed to the additional IREE due to the strong ISOC between Co - 3$d$ and Pt - 5$d$ orbitals at the Pt/Co interface via bringing in an ultrathin Co interlayer, which is confirmed by ISOC-related spin properties characteristics, such as a large IPMA energy density $K_s = 2.7$ erg/cm$^2$, a strong TMS coefficient $\beta_{TMS} = 0.46$ nm$^2$ and a moderate effective $G_{eff,tot}^{\uparrow\downarrow} = 0.42 \times 10^{15} \Omega^{-1} m^{-2}$. Additionally, a considerable $\sigma_z$-spin currents with a maximum at $t_{Co} = 0.4$ nm, which are proportional to the $H_\perp$, suggesting that the generated $\sigma_z$-spin currents are correlated to the effective $M_z$ component. Our results suggest that the interfacial magnetic properties, spin current generation, and spin transport in the hybrid heterostructure thin film can be effectively manipulated by tailoring their interfaces.

**Acknowledgments:** The project is supported by the National Natural Science Foundation of China (Grants No. 11774150, 12074178, 11874135, and 12004171), the Applied Basic Research Programs of the Science and Technology Commission Foundation of Jiangsu Province (Grant No. BK20200309), the Open Research Fund of Jiangsu Provincial Key Laboratory for Nanotechnology, Key Research and Development Program of Zhejiang Province under Grant No. 2021C01039, and the Scientific Foundation of Nanjing University of Posts and Telecommunications (NUPTSF) (Grant No. NY220164).